\documentstyle[aps,prb,floats]{revtex}
\input{psfig}
\begin{document}
\draft
\twocolumn[\hsize\textwidth\columnwidth\hsize\csname@twocolumnfalse\endcsname

\title{A quantitative theory of current-induced step bunching on Si(111)}
\author{Da-Jiang Liu$^1$ and John D. Weeks$^{1,2}$}
\address{$^1$Institute for Physical Science and Technology and\\
$^2$Department of Chemistry,\\
University of Maryland, College Park, Maryland 20742}
\date{\today}
\maketitle

\begin{abstract}
We use a one-dimensional step model to study quantitatively the growth of
step bunches on Si(111) surfaces induced by a direct heating current.
Parameters in the model are fixed from experimental measurements near 900$%
^\circ$C under the assumption that there is local mass transport through
surface diffusion and that step motion is limited by the attachment rate of
adatoms to step edges. The direct heating current is treated as an external
driving force acting on each adatom. Numerical calculations show both
qualitative and quantitative agreement with experiment. A force in the step
down direction will destabilize the uniform step train towards step
bunching. The average size of the step bunches grows with electromigration
time $t$ as $t^\beta $, with $\beta \approx 0.5$, in agreement with
experiment and with an analytical treatment of the steady states. The model
is extended to include the effect of direct hopping of adatoms between
different terraces. Monte-Carlo simulations of a solid-on-solid model, using
physically motivated assumptions about the dynamics of surface diffusion and
attachment at step edges, are carried out to study two dimensional features
that are left out of the present step model and to test its validity. These
simulations give much better agreement with experiment than previous work.
We find a new step bending instability when the driving force is along the
step edge direction. This instability causes the formation of step bunches
and antisteps that is similar to that observed in experiment.
\end{abstract}

\pacs{68.35.Ja,68.10.Jy,68.55.Jk,05.70.Ln}

]

\section{Introduction}

In 1989 Latyshev {\em et al.}\cite{latyshev} made the startling discovery
that a direct heating current can induce step bunching on vicinal Si(111)
surfaces. When the sample is resistively heated with direct current, steps
can rearrange into closely spaced step bunches separated by wide terraces.
Around 900$^\circ$C, the step train is unstable towards step bunching when
the current is in the step-down direction, but is stable when the current
direction is reversed. Surprisingly, as the temperature is increased to 1190$%
^\circ$, the stable and unstable current directions are reversed, i.e., the
step train is unstable with step-up current and stable with step-down
current. There is another such reversal as the temperature is increased
further.

Since then the phenomenon has received a great deal of attention.
Theoretical work has mainly concentrated on two goals: understanding the
microscopic physics underlying the instability towards step bunching and the
reversal of the unstable current direction with temperature, \cite
{stoyanov,kandel96,misbah} and determining the mesoscopic evolution of the
surface morphology as a result of the instability.\cite
{kandel95,sato,natori,misbah96,krug,krug95,dobbs96} Recently Williams {\em %
et al.}\cite{williams4,yang1,fu,fu97} carried out a series of measurements
on Si(111) surfaces at 900$^{\circ }$C, to provide a quantitative
understanding of the dynamics. By controlling the experimental system and
comparing with theoretical models, they were able to extract detailed
information about the mechanism and to determine quantitative values of
relevant parameters. Although the details of the microscopic mechanisms
leading to the change in the destabilizing current direction with varying
temperature are still not fully understood,\cite{stoyanov,kandel96,misbah}
we show here that there exists a reliable mesoscopic theory that can provide
quantitative agreement with a variety of experimental results in the
temperature regime (900$^{\circ }$C) studied by Williams {\em et al.}

In Secs.~\ref{sec:model} and \ref{sec:experi}, we briefly review some of the
experimental and theoretical work that led to our present model. We focus on
the case where the {\em step motion is limited by the attachment rate of
adatoms to the step edge} (in contrast to being limited by the diffusion
rate on terraces). We also assume {\em local mass transport} by surface
diffusion. These assumptions yield a minimal mesoscopic model that is
consistent with all previous experimental results. In Sec.~\ref{sec:results}
we give numerical results from this model using realistic parameter values
and interpret and analyze some of the results in Sec.~\ref{sec:discussion}.
We briefly discuss in Sec.~\ref{sec:perm} some effects of step permeability 
\cite{chalmers,tanaka97} (direct adatom hops from one terrace to another),
which might be important in other systems, e.g., Si(001). In Sec.~\ref{MC}
we present some results of Monte-Carlo simulations of a microscopic
solid-on-solid model, using physically motivated assumptions about the
dynamics of surface diffusion and attachment at step edges. These results
are in qualitative agreement with experiment, in contrast to previous work 
\cite{krug,krug95,dobbs96} using conventional Metropolis dynamics. They also
help in the understanding of additional 2D features and instabilities that
cannot be described by the simple 1D step model. Final remarks are given in
Sec.~\ref{sec:conclusion}.

\section{1D step model with extended velocity functions}

\label{sec:model}

Vicinal surfaces, which are created by a miscut along a low-index plane
below the roughening temperature, are most naturally and usefully described
by a model of interacting steps of the same sign. Because of the inherent
anisotropy of the underlying crystal structure, these surfaces often exhibit
quasi one dimensional features, thus making a 1D step model useful and
accurate. Also the number of steps is often conserved as the surface
evolves, permitting further simplifications in the analysis.

The change in the morphology of vicinal surfaces can be described in terms
of the velocity of each step as long as no steps are created or destroyed.
The classic Burton-Cabrera-Frank (BCF) \cite{burton} treatment assumes that
the mass transfer is governed by a set of adatom diffusion equations on each
terrace, with steps acting as perfect sinks and sources for adatoms (or
vacancies) so that local equilibrium is always maintained. However, the
original BCF picture is valid only for simple materials where adatom
diffusion is the rate limiting process. Extensions of the BCF model can be
made to include a finite attachment/detachment term in the boundary
conditions. This is needed for materials like the silicon, where the atom
exchange rate between steps and terraces is not fast enough to permit the
adatom concentration near the step edge to achieve local equilibrium.

In his important work on the instability induced by a direct heating current
on Si(111), Stoyanov \cite{stoyanov} proposed such a modified BCF model,
including both a finite adatom attachment/detachment rate at step edges and
an adatom drift velocity (or equivalently, an external driving force due to
the electric field). Natori \cite{natori} extended the work of Stoyanov to
include step repulsions. The idea of incorporating step interactions in a
generalized BCF model has been further developed by Sato and Uwaha.\cite
{sato}

\begin{figure}[t]
\centerline{\psfig{file=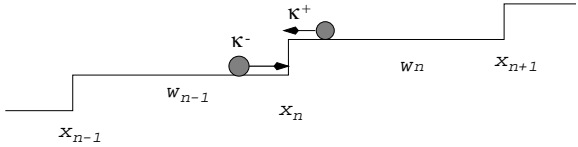,width=3in}}
\caption{Illustration of the labeling of steps and terraces and kinetic
coefficients.}
\label{notations}
\end{figure}

To describe the exchange of atoms or vacancies between steps and their
neighboring terraces (attachment/detachment), we use a linear kinetics
theory and write the net surface flux from step $n$ to terrace $n$ (see Fig. 
\ref{notations} for the labeling) as 
\begin{mathletters}
\begin{equation}
j_n^{+}=\frac{\kappa _{+}c_{{\rm eq}}}{kT}[\mu _n-\mu ^t(x_n^{+})],
\label{jnp}
\end{equation}
and the flux from step $n$ to terrace $n-1$ as 
\begin{equation}
-j_n^{-}=\frac{\kappa _{-}c_{{\rm eq}}}{kT}[\mu _n-\mu ^t(x_n^{-})],
\label{jnm}
\end{equation}
where $\mu _n$ is the atom chemical potential at step $n$, defined as the
increase in free energy per atom when atoms attach to the step. For a 1D
step train with elastic and entropic step repulsions, this can be written as 
\cite{rettori} 
\end{mathletters}
\begin{equation}
\mu _n=2gh^3a^2\left( \frac 1{w_{n-1}^3}-\frac 1{w_n^3}\right) ,  \label{mu}
\end{equation}
where $a^2$ is the area of a single atomic cell on the surface, $h$ the
single step height, and $w_n\equiv x_{n+1}-x_n$ the width of terrace $n$.
Here the parameter $g$ is just the coefficient of the $s^3$ term in the
well-known Gruber-Mullins \cite{gruber} form for the projected free energy
of vicinal surfaces with slope $s$. $\mu ^t(x_n^{\pm })$ is the adatom
chemical potential on the terrace adjacent to step $n$, approaching $x_n$
from right hand ($+$) and left hand ($-$) side.\cite{nozieres1a} $c_{{\rm eq}%
}$ is the equilibrium adatom concentration on terraces. We assume there is
no asymmetry for the kinetic coefficient for adatom attachment from upper
and lower terraces ($\kappa _{+}=\kappa _{-}$).

To determine the adatom chemical potential on terraces $\mu ^t(x)$, we need
to know the mass transport mechanism on the surface. If it is much easier
for adatoms to hop directly across a step edge from one terrace to another
one in comparison to attachment at the step edge, then the adatom chemical
potential becomes a constant on all terraces. We refer to this as {\em %
non-local mass transport} (case A). The other limit is when there is no
significant hopping of adatoms over the step edge, as assumed in the BCF
model; we call this the limit of {\em local mass transport} (case B).
Experiments on Si(111) show that the relaxation rate of a step bunch with $N$
steps scales with $N^{-\alpha }$ where $\alpha =4.3 \pm 0.5$. As we have
discussed in detail elsewhere,\cite{fu,liu96} this is consistent with the 
{\em local mass transport limit} (case B), and we will assume this limit in
most of this paper. In Sec. \ref{sec:perm} we will consider a more general
scenario.

Assuming local mass transport, the step velocities $v_n(t)$ can be
determined by solving the diffusion equation for adatoms on terraces with
boundary conditions at step edges governed by linear kinetics. The equations
can be written generally in an {\em extended velocity function form}:\cite
{liu97a,weeks97} 
\begin{equation}
v_n=f_{+}(w_n;\mu _n,\mu _{n+1})+f_{-}(w_{n-1};\mu _{n-1},\mu _n)
\label{vel}
\end{equation}
We will not write down the general form for the velocity functions $f_{\pm }$
for the electromigration problem since it is very complicated and not very
instructive for our purpose. A simple limit that is consistent with
experiment will be discussed below. More general expressions have been given
by many authors.\cite{note}

In studies of surface dynamics, it is often convenient to consider idealized
models where the kinetics is limited by a few slow processes on the surface,
and the rates of other faster processes are taken to infinity. For BCF
models, neglecting evaporation and deposition, there are two basic rates,
the attachment/detachment rate $\kappa $, and the adatom diffusion rate $D_s$%
. Bartelt {\em et al.} \cite{bartelt} estimated the attachment/detachment
rate from the step fluctuations of Si(111) at 900$^{\circ }$C under the
assumption that attachment/detachment is the rate limiting process, while
Pimpinelli {\em et al.} \cite{pimpinelli1} estimated from the same data the
diffusion rate under the assumption that adatom diffusion is the rate
limiting process. It is useful to define a length scale $d\equiv D_s/\kappa $%
. When $d$ is very small, the step dynamics is said to be {\em diffusion
limited}, and when $d$ is very large, the dynamics is {\em %
attachment/detachment limited}.

\begin{table}[t]
\caption{Sets of parameters which give a good fit to the relaxation of step
bunches. We use $\Gamma =2c_{{\rm eq}}a^4\kappa $ ($c_{{\rm eq}}a^2=0.2{\rm %
ML}$) to compare with previous work. The table is taken from Ref. 15.}%
\begin{tabular}{cdddl}
Parameter Set & $d$(\AA) & $q (e)$ & $\Gamma ({\rm \AA^3/s})$ &
$D_s({\rm \AA^2/s}$) \\ 
A & 100000 & 0.006 & $3 \times 10^7$ & 5.2 $\times 10^{11}$ \\
B & $5000$ & 0.006  & $4 \times 10^7$ & 3.4 $\times 10^{10}$ \\
C & $100$ & 0.03 & $3\times 10^8$ & 5.2 $\times 10^9$ \\
D & $10$ & 0.2 & $2\times 10^9$ & 3.5 $\times 10^8$ 
\end{tabular}
\end{table}

However, direct estimation of this ratio is difficult. For example, Table I
lists several sets of parameters that give good
agreement with experiments \cite{fu97} on the electromigration-driven
relaxation of step bunches on Si(111), with $d$ ranging from $10{\rm \AA }$
to $10^5{\rm \AA }$. Physically $d$ has to be finite and whether a system is
diffusion limited or attachment/detachment limited depends on the comparison
of $d$ with other length scales, e.g., the typical terrace width. Here we
refer to the mathematical limit $d\rightarrow \infty $ as the {\em complete
attachment/detachment limited} model, and we call the limit $d\rightarrow 0$
the {\em complete diffusion limited} model.

As suggested by Table I, the relaxation experiments can be explained using
an effective charge $q$ of the adatoms that reaches a finite and physically
reasonable value as $d\rightarrow \infty $. In contrast, $q$ must tend to
the unphysical limit $\infty $ when $d\rightarrow 0$. Therefore the complete
attachment/detachment limited model is well-defined, and we will use this
limit to illustrate the mechanism for electromigration. As we show below,
Eq.~(\ref{vel}) simplifies considerably in this limit. Moreover, a value of $%
d\ge 3000{\rm \AA }$ is predicted by extrapolating the diffusion rate from
higher temperatures using a diffusion activation energy \cite{yang1} of
1.1eV. However, numerical solutions of Eq.~(\ref{vel}) using any of the
parameter sets in Table I are consistent with the step bunching experiments,
including the power law for coarsening.

As in the Stoyanov model,\cite{stoyanov} we assume that there is a force $F$
acting on each adatom because of the electric field. The adatom flux on
terrace $n$ under this driving force is 
\begin{equation}
j_n=\frac{D_sc_{{\rm eq}}}{kT}\left( -\frac{\partial \mu ^t}{\partial x}%
+F\right) ,  \label{jn}
\end{equation}
With complete attachment/detachment limited kinetics, $D_s$ tends to
infinity relative to the attachment rate $\kappa $, or $j_n$. Therefore the
adatom chemical potential $\mu ^t(x)$ has a {\em constant gradient}, $F$, on
each terrace. In general, $\mu ^t(x)$ is affected by the motion of the
neighboring steps, but usually the steps move very slowly so that they can
be treated as effectively stationary as far as the diffusion of adatoms is
concerned. Under this {\em quasistatic approximation},\cite{quasistatic} at
any given time, the total surface flux into terrace $n$ from the two
neighboring steps $(j_n^{+}-j_{n+1}^{-})$ equals to the total amount of
evaporation from this terrace, which is given by $c_{{\rm eq}}w_n/\tau
\equiv w_n/\tau _e$, where $\tau $ is the average lifetime of an adatom on
the terrace before it evaporates. With these approximations, the step
velocities in Eq.~(\ref{vel}) can be written in the simple form \cite
{liu97a,weeks97} 
\begin{equation}
v_n=\frac{\kappa c_{{\rm eq}}a^2}{2kT}(2\,\mu _n-\mu _{n+1}-\mu
_{n-1})+k_{+}w_n+k_{-}w_{n-1},  \label{linearvel}
\end{equation}
where 
\begin{equation}
k_{\pm }=\pm \frac{\kappa c_{{\rm eq}}a^2F}{2kT}+\frac 1{2\tau _e}.
\end{equation}

\section{Experimental determination of the parameters}

\label{sec:experi}

Williams {\em et al.} devised a series of experiments to measure the various
parameters and also test the assumptions about the mass transport limits
discussed in the previous section. As mentioned earlier, on Si(111) at $%
900^{\circ }$C, the relaxation of step bunches is consistent with the local
mass transport limit. The step interaction parameter $g$ can be measured 
\cite{alfonso,williams2} from the distribution of terrace widths and step
positions at equilibrium, and has an estimated value around $0.015{\rm %
eV/\AA ^2}$. Assuming attachment/detachment limited kinetics for mass
transport, the kinetic coefficient $\kappa $ can be measured independently
from the thermal fluctuations of the steps and the relaxation of step
bunches. Rather than using $\kappa $, we define $\Gamma =2c_{{\rm eq}%
}a^4\kappa $ to compare with earlier work. $\Gamma $ gives the step mobility
for the Brownian motion of an isolated step \cite{bartelt} and is measured
to be around $5\times 10^7{\rm \AA ^3/s}$. This value also gives a good fit
to the relaxation of step bunches, thus providing additional evidence
supporting the local mass transport assumption aside from the scaling
behavior mentioned before.

The force on adatoms due to the direct heating current can be measured from
the relaxation of the step bunches that occurs after reversing the current
to the stable direction.\cite{fu97} The force acting on each adatom can be
conveniently described in terms of an effective charge $q,$ with $F=qE$,
where the experimental value of $E=7{\rm V/cm}$. Table I lists four sets of
parameters that give good fits to the decay of step bunches with a direct
current in the stabilizing direction at 900$^{\circ }$C. As mentioned
before, as $d$ becomes very large, the values of $q$ and $\Gamma $ reach
limiting values, which we use in the complete attachment/detachment limited
model. Other relevant parameters include the average terrace width $w_0=1100%
{\rm \AA }$, and the evaporation time for one monolayer $\tau _e=1250{\rm s}$%
.\cite{yang1}

\section{Numerical results and comparison with experiments}

\label{sec:results}

\begin{figure}[t]
\centerline{\psfig{file=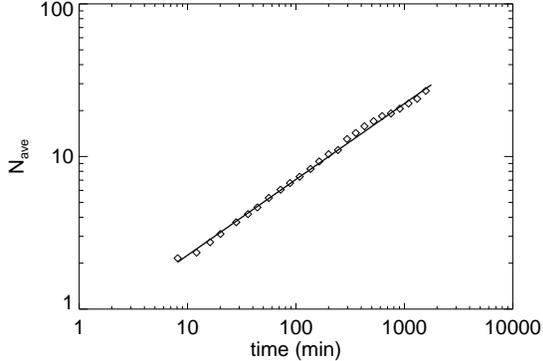,width=3.in}}
\caption{Evolution of average bunch size using parameter set A. The solid
line is a fit to $t^\beta$ with $\beta = 0.50$.}
\label{nave}
\end{figure}

When the adatom drift velocity is in the step down direction ($F<0$), one
can show from a linear stability analysis using the parameters in Table I
that a uniform step train is unstable towards step bunching. The evolution
of the step bunches is determined by numerically integrating Eq.~(\ref{vel}%
), starting from a step train with small random deviations from uniform
spacing. The system continues to coarsen by forming larger and larger step
bunches. Figure \ref{nave} shows the average bunch size $N_{{\rm ave}}$ for
a system of 2049 steps as a function of time, using Eq.~(\ref{linearvel})
for the complete attachment/detachment limited model. A bunch is defined by
a number of adjacent steps with no terraces between them larger than $w_0/2$%
. The average bunch size is defined by $\sum_nn\rho _n/\sum_n\rho _n$, where 
$\rho _n$ is the number (or density) of bunches consisting of $n$ step. It
can be fitted by a $t^\beta $ growth rate with $\beta \approx 0.50$. Results
using the more complicated velocity functions in Eq. (\ref{vel}) obtained
from solutions of a generalized BCF equation and parameter set $A$ are
almost indistinguishable on this scale.

This compares very well with the STM results of Yang, Fu, and Williams.\cite
{yang1} They show at both 945$^{\circ }$C and 1245$^{\circ }$C that the
growth of the facet sizes $Z$ between two step bunches satisfies $t^\beta $
where $\beta \approx 0.5$. It is a good approximation to relate the average
number of steps in a bunch $N_{{\rm ave}}$ to $Z/w_0$. They observed that at
945$^{\circ }$C, after about 120 min of annealing time, the average terrace
width between step bunches grows to about $3500{\rm \AA }$. In the numerical
simulations, the average terrace width grows to about $6800{\rm \AA }$ in
the same time. We consider this quite satisfactory agreement, given the
uncertainties in the values of the experimental parameters we used. Thus,
not only does the step model give the correct power law growth rate, it also
gives good quantitative agreement with experimental results at $945^{\circ }$%
C.

Simulations using other sets of parameters in Table I produce slightly
different results, but all agree with the experimental data within the
errors in measured parameters. Moreover, they all have approximately a $%
t^{1/2}$ coarsening rate during the time simulated and experimentally
observed. Therefore we can not determine a unique set of microscopic
parameters accurately from the coarsening rate alone.

Dobbs and Krug \cite{dobbs96} also obtained a $t^{1/2}$ coarsening rate from
simulations of a 2D solid-on-solid model using Metropolis dynamics. However,
they obtained the $t^{1/2}$ behavior only when there is significant {\em %
lateral} fluctuations of step bunches as can sometimes occur in the later
stages of coarsening, while initially the growth law they observed went as
$t^{1/4}$. Experimentally there is no such transition, and we obtain this kind of
coarsening from a 1D model with straight steps. Moreover, as discussed below
in Sec. \ref{MC}, there are other unphysical surface features that arise
from the use of Metropolis dynamics to describe Si(111) and related systems,
and we suggest there an alternative dynamical scheme that gives good
qualitative agreement with experiment.

\begin{figure}[t]
\centerline{\psfig{file=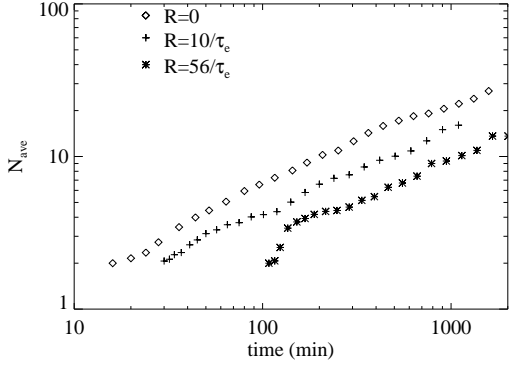,width=3.in}}
\caption{Effects of growth conditions on the step bunching.}
\label{growth}
\end{figure}

As another application of the step model, we also simulate the step bunching
occurring under {\em growth conditions}. It is well known that a Schwoebel
barrier \cite{schwoebel69} has very different effects on growth and
evaporation. For example, if there is an additional barrier for an adatom to
attach to a step edge from the upper terrace, a 1D step train will be
stabilized under growth and destabilized under evaporation. However, even in
the absence of a Schwoebel asymmetry (that is, even when $\kappa _{+}=\kappa
_{-}$) as assumed here, simulations of the present step model under growth
conditions show a {\em decrease} in the bunching rate with increasing
deposition rate. Figure \ref{growth} shows the dependence of the average
bunch size as a function of time for different growth conditions. It is
useful to define $\tilde{R}=R\tau /c_{{\rm eq}}$ as the ratio between
deposition and desorption rates. As $\tilde{R}$ increases, the bunching rate
decreases, in good agreement with the experimental results of Yang, Fu, and
Williams.\cite{yang1} This decrease in the coarsening rate with increasing
deposition was also noted by Tersoff et al. \cite{tersoff} in their study of
stress induced step bunching. As Kandel and Weeks \cite{kandel95} argued,
when the step train is traveling in a certain direction (e.g., due to
deposition or evaporation), a step at the front end of a step bunch can
leave the bunch and join with the step bunch in front of it, causing an
exchange of steps between step bunches. As the growth rate increases, the
velocities of these crossing steps get larger and larger, so more debunching
occurs, thus reducing the coarsening rate. A detailed study of the effect of
debunching requires a 2D model and is beyond the scope of this paper.

\section{Analysis: quasi-steady state and coarsening rate}

\label{sec:discussion}

In this section we will try to understand analytically some of the numerical
results. Although it is straightforward to use the solution of the diffusion
equation to determine the velocity functions in Eq.~(\ref{vel}) when
simulating the step bunching numerically, it is more convenient and
instructive to consider the simple linear velocity function model of Eq.~(%
\ref{linearvel}). Initially we also neglect any deposition or evaporation.

\begin{figure}[t]
\centerline{\psfig{file=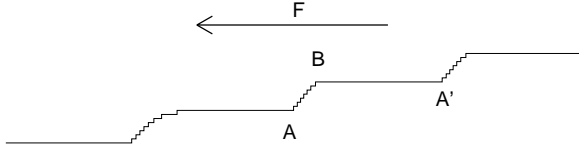,width=3.in}}
\caption{Typical profile of step configurations when step bunches are
induced by a force acting on adatoms in the step down direction.}
\label{stepprofile}
\end{figure}

When the surface flux $j$ is a constant everywhere, the surface is in a
steady state. As we will show later, such a steady state is possible when $F$
is in the step down direction. Figure \ref{stepprofile} is snapshot of the
profile of the surface during one particular simulation. Assuming that the
step bunch between $A$ and $B$ is in a steady state with a surface atom flux 
$j^{*}$, we have from Eqs.~(\ref{jnp}) and (\ref{jnm}), 
\begin{equation}
(N_s-1)j^{*}=\frac{\kappa c_{{\rm eq}}}{2kT}(\mu _A-\mu _B+F\,w_{AB}),
\end{equation}
where $N_s$ is the number of steps in the bunch and $w_{AB}$ is the distance
between $A$ and $B$. On the flat terrace between $B$ and $A^{\prime }$ the
flux is given by 
\begin{equation}
j^{*}=\frac{\kappa c_{{\rm eq}}}{2kT}(\mu _B-\mu _{A^{\prime
}}+F\,w_{BA^{\prime }}).
\end{equation}
For a {\em periodic array} of step bunches, $\mu _A=\mu _{A^{\prime }}$, and
the steady state flux is then 
\begin{equation}
j_{{\rm ss}}^{*}=\frac 1{N_s}\frac{\kappa c_{{\rm eq}}}{2kT}%
\,F\,w_{AA^{\prime }}=\frac{\kappa c_{{\rm eq}}\,F\,w_0}{2kT}.  \label{jstar}
\end{equation}
Therefore the steady state adatom current is independent of the size of the
step bunches. This result is valid for a periodic array of step bunches and
is important in determining the time scaling exponent, as it will be shown
at the end of this section.

The steady state profile of the step bunch can be calculated numerically.
Here we go to the continuum limit, which is a good approximation when the
number of steps in the bunch is large. If $z(x)$ denotes the height of the
surface, then the slope is $z_x(x)$ $\equiv \partial z/\partial x$. Using
the continuum version of the Gruber-Mullins free energy functional \cite
{gruber,nozieres} 
\begin{equation}
H=L_y\int g|z_x(x)|^3\,dx,
\end{equation}
which is appropriate for vicinal surfaces below the terrace roughening
temperature, we can write the adatom chemical potential as \cite{nozieres} 
\begin{equation}
\mu (x)=a^2h\frac{\delta H}{\delta z}=-6ga^2h|z_x|z_{xx}.  \label{mucont}
\end{equation}
Here $z_{xx}$ is the second derivative of $z(x)$ with respect to $x$. As in
the step model discussed earlier, we can drop the linear step energy term if
no new steps are created. Note that in the continuum description slopes of
different signs correspond to positive and negative steps. Here we consider
step profiles with positive slopes everywhere and thus can set $|z_x| = z_x$%
. In the attachment/detachment limit, the surface flux is given by 
\begin{equation}
j(x)=\frac{\kappa c_{{\rm eq}}}{2\,kT}\frac h{z_x}\left[ F-\frac \partial {
\partial x}\mu (x)\right] .  \label{jxF}
\end{equation}

The above equation has the physical property that the adatom mobility (as
the response to an external field) is inversely proportional to the slope $%
z_x$. To calculate the step bunch profile $z(x$), we can neglect the first
term in Eq. (\ref{jxF}) if the terrace width inside a bunch is much smaller
than the distance between bunches. For an isolated step bunch with $%
j(x)=j^{*}$, using Eq.~(\ref{mucont}), we have 
\begin{equation}
\frac{3gh^2\kappa c_{{\rm eq}}}{kT}\frac \partial {\partial x}%
(z_xz_{xx})=j^{*}z_x.
\end{equation}
This can be reduced to 
\begin{equation}
z_x^3=\frac{j^{*}kT}{2\kappa gh^2c_{{\rm eq}}}\left[
(z-z_0)^2-(H/2)^2\right] ,  \label{profile}
\end{equation}
where $z_0$ and $H$ are integration constants. The step bunch profile $z(x)$
can be easily calculated numerically by integrating Eq.~(\ref{profile}), or
analytically in terms of hypergeometric functions. Equation (\ref{profile})
was first derived by Nozi\`{e}res \cite{nozieres} in his study of surface
dynamics below the roughening transition. The maximum slope of the step
bunch with $j^{*}<0$ is given by $z_x^{{\rm max}}=[-j^{*}(H/2)^2kT/(2\kappa
gh^2c_{{\rm eq}})]^{1/3}$. For a true steady state with a periodic array of
step bunches, $j^{*}=j_{{\rm ss}}^{*}$ [Eq.~(\ref{jstar})]. We expect $j^{*}$
to fluctuate around this value for a system in a quasi steady state. $H$ is
approximately the height of the step bunch for large bunches. We have $z_x^{%
{\rm max}}\sim N_s^{2/3}$ since $j_{{\rm ss}}^{*}$ is independent of $N_s$.
This can be experimentally tested by measuring the average slope of the step
bunch as a function of the average bunch size. Note that the continuum limit
breaks down near the edges of a step bunch where sharp changes in the local
slope occur.

Strictly speaking, the above analysis only holds for steady state profiles
with complete attachment/detachment limited kinetics, but we expect it to be
a good approximation for the quasi steady state profiles that arise as the
step bunches slowly coarsen with time. Indeed simulations and experiment
agree that the step bunches coarsen with time as $t^\beta $ with $\beta
\approx 0.5$. This $t^{1/2}$ power law can be justified by a scaling
argument. We assume that as $t\rightarrow \infty $, there is only one
characteristic length for the system, which scales as $t^\beta $. We can
thus write all the variables on the surface in term of the scaled length $%
x/t^\beta $ at time $t$. We have noted that the steady state flux is
independent of the size of the step bunches [Eq.~(\ref{jstar})]. For a
system in a quasi-steady state, we thus assume that the flux can be written
as a function of the scaled length only with no extra time dependence, i.e., 
\begin{equation}
j(x,t)=J(x/t^\beta ).
\end{equation}
In contrast, the surface profile should maintain a constant average slope
and thus must have the following scaling form: 
\begin{equation}
z(x,t)=t^\beta Z(x/t^\beta ).
\end{equation}
Substituting these into the equation expressing microscopic mass
conservation: 
\begin{equation}
\frac \partial {\partial t}z(x,t)\sim -\frac \partial {\partial x}j(x,t),
\end{equation}
we have $\beta =1/2$ by comparing the leading exponents on both sides of
this equation. This prediction is in good agreement with both the
experimental and the numerical work, as shown in Fig.~\ref{nave}.

\section{Effects of step permeability}

\label{sec:perm}

In the previous sections we focused our study of the step model on vicinal
Si(111) surfaces around 900$^{\circ }$C. Our basic approach can be applied
more generally, though the limits we used above are not necessarily
satisfied. However as long as the mass transport is local, other differences
from our present model, e.g., a finite diffusion length or asymmetric step
edge attachment rates (Schwoebel barriers), can be studied using the
extended BCF model and the velocity function approach of Eq.~(\ref{vel}) in
a more or less straightforward way.

We will not detail this work here, but instead turn to the conceptually
interesting case of {\em step permeability}, which can make the extended BCF
picture no longer valid. This is motivated in part by recent work by Tanaka 
{\em et al.} \cite{tanaka97} and Stoyanov.\cite{stoyanov2} In the classic
BCF picture with local mass transport, the only way to achieve adatom
transport from one terrace to another one is through attachment to and
subsequent detachment from the step edge separating them. However, if there
is {\em direct hopping} of adatoms across a step edge without incorporating
into the step edge first, this causes a coupling of diffusion fields on
adjacent terraces that must be taken into account.

In a BCF-like picture, the adatom chemical potentials can have
discontinuities at step edges either because steps are perfect sinks, or
because there is a strong diffusion barrier near the step edge. In their
analyses of island flattening on Si(001), Tanaka {\em et al.} \cite{tanaka97}
introduced an adatom hopping term between adjacent terraces over a step edge
proportional to the difference between the chemical potentials on the two
terraces. Although this inter-terrace hopping could be fast compared with
attachment of adatoms to the step edge, it could still be slow compared with
diffusion on flat terraces, thus allowing discontinuities in the adatom
concentration field at step edge positions. Here we take this limit,
assuming that the adatom diffusion rate on flat terraces is always much
faster than {\em both} the attachment and inter-terrace hopping rates.

\begin{figure}[t]
\centerline{\psfig{file=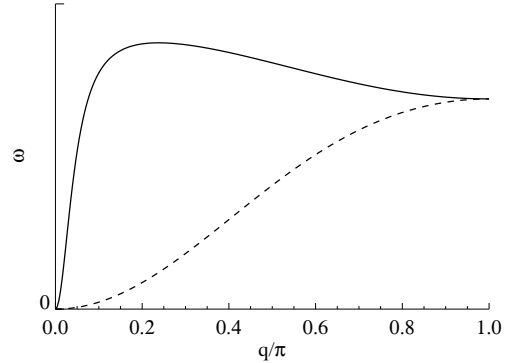,width=3.in}}
\caption{Linear instability of a uniform step train. The local hopping of
adatoms over step edges (step permeability), coupled with the step
repulsions, moves the maximum instability away from the step pairing mode ($
q=\pi $) to longer wavelengths. The solid line is for $p/\kappa =100$ and
the dashed line is for $p=0$. Other parameters are taken from set A in Table
I, although we don't attempt to describe Si(111) realistically here.}
\label{omega}
\end{figure}

The effect of step permeability on the relaxation of step bunches due to
step repulsions can be studied straightforwardly. In the absence of an
external driving force, the adatom chemical potential on each terrace is a
constant denoted by $\mu _n^t$ (see Fig.~\ref{notations} for the labeling of
terraces). We can write the net surface flux at the right hand side of step $%
n$ as 
\begin{mathletters}
\begin{equation}
j_n^{+}\sim \kappa \,(\mu _n-\mu _n^t)+p\,(\mu _{n-1}^t-\mu _n^t),
\label{permplus}
\end{equation}
and the flux at the left hand side of step $n$ as 
\begin{equation}
j_n^{-}\sim \kappa \,(\mu _{n-1}^t-\mu _n)+p\,(\mu _{n-1}^t-\mu _n^t).
\label{permminus}
\end{equation}
Assuming again the quasistatic limit where $j_n^{+}=j_{n+1}^{-}$, we can
solve $\mu _n^t$ for any given set of $\mu _n$ from a system of linear
equations. The analytic solution is given in the Appendix. It is easy to see
that the $p=\infty $ and $p=0$ limits correspond to case A and case B
dynamics respectively.

As was mentioned in Sec. \ref{sec:model}, experiments \cite{fu} on the
relaxation rate of step bunches on Si(111) near 900$^{\circ }$C show a size
scaling exponent $\alpha = 4.3 \pm 0.5$, consistent with the $p=0$ limit ($%
\alpha=4$). In comparison, if we assume $p=2\kappa $, we obtain $\alpha =3.6$
for the bunch sizes used in experiments and larger $p$ will give even
smaller $\alpha $. Therefore we conclude $p<2\kappa $ and can dismiss the
importance of step permeability for Si(111) at 900$^{\circ }$C. However $p$
can be large for other systems, or even perhaps for Si(111) at different
temperatures. Tanaka {\em et al.} \cite{tanaka97} estimated $p=36\kappa $ on
Si(001) at 950$^{\circ }$C. Here we discuss some interesting effects step
permeability has on electromigration, which could be a way to detect any
significant permeability if it exists.

The adatom concentration field has a constant gradient on each terrace when
there is a driving force. We need to be more precise in our description of
the microscopic origin of step permeability to obtain a complete theoretical
description. Here we consider the case where the step permeability is
proportional to the difference between the {\em local} adatom chemical
potentials immediately to the left and right hand sides of the step.
Equations (\ref{permplus}) and (\ref{permminus}) then become 
\end{mathletters}
\begin{mathletters}
\begin{equation}
j_n^{+}\sim \kappa \,[\mu _n-\mu ^t(x_n^{+})]+p\,[\mu ^t(x_n^{-})-\mu
_n^t(x_n^{+})],
\end{equation}
and 
\begin{equation}
j_n^{-}\sim \kappa \,[\mu ^t(x_n^{-})-\mu _n]+p\,[\mu ^t(x_n^{-})-\mu
_n^t(x_n^{+})].
\end{equation}
$\mu ^t(x)$ can be determined in much the same way as before by assuming
there is a gradient, $F$, in $\mu (x)$ on each individual terrace. In a
uniform step train, the external force creates a local chemical potential
gradient, and introduces a ``leak'' of surface flux from the permeability
term. We now show that the ``leak'' will create a {\em long wavelength}
bunching instability, in contrast to the pairing instability familiar from
the BCF picture.

In a linear stability analysis, the step positions are written as 
\end{mathletters}
\begin{equation}
x_n(t)=\sum_\phi e^{in\phi +\omega (\phi )t}u_\phi (0)+n\,w_0,
\end{equation}
where 
\begin{equation}
u_\phi (0)=\frac 1N\sum_ne^{-in\phi }[x_n(0)-n\,w_0],
\end{equation}
for small perturbations from uniform configurations. Figure \ref{omega}
plots the amplification exponent $\omega $ of a uniform step train as a
function of a dimensionless wavenumber ($\phi =\pi $ corresponds to the
pairing mode). The solid line is for $p/\kappa =100$ and the dashed line is
for $p=0$. The maximum linear instability has shifted to much longer
wavelengths. Note that very strong repulsive interactions could also produce
such a shift.\cite{sato} However, for systems with step permeability, there
is a very rapid (almost linear) growth in the average size of step bunches
in the initial stage, which then crosses over to the $t^{1/2}$ behavior. In
contrast, for the purely repulsive system, the growth rate is approximately $%
t^{1/2}$ at all times. These characteristics could be used to detect step
permeability if it is very large.

\section{Monte-Carlo simulations of the 2D solid-on-solid model: modified
Arrhenius dynamics}

\label{MC}

For Si(111) at 900$^{\circ }$C the steps are mostly straight and a 1D model
is adequate for most purposes. However, at higher temperatures, there exists
noticeable bending of steps. For example, at around 1100$^{\circ }$C when
evaporation is significant, 2D arrays of {\em crossing steps }form between
step bunches. Kandel and Weeks \cite{kandel95} proposed a (quasi) 2D step
model where the velocity of each step depends only on the {\em local}
neighboring terrace widths in the direction perpendicular to the average step
edge direction. This model reproduced many features of the crossing arrays
quite accurately.\cite{williams4} Further developments along these lines
have been reported in Refs.~\onlinecite{liu97a} and \onlinecite{weeks97}.

A full 2D step model taking account of 2D adatom diffusion on terraces with
boundary conditions on the moving curved steps is very difficult to study.
Also it is necessary to go beyond the BCF framework which excludes the
creation of new steps to explain the anti-step bunches reported by Latyshev 
{\em et al.} \cite{latyshev94} Here we study a generalized 2D solid-on-solid
(SOS) model that takes explicit account of a {\em step edge barrier} in the
kinetics of adatom attachment/detachment at step edges. We believe this is
probably the simplest 2D microscopic model that can provide a physically
reasonable description of both adatom diffusion and step motion in Si(111)
and related systems. However there is almost no hope of simulating the long
time behavior of such a microscopic model using realistic parameter values.
Thus, in contrast to the 1D step model we studied above, here we concentrate
only on qualitative properties. Specifically, we consider a very large
external driving force along with a very small average terrace width. These
extreme choices will permit significant step motion in the computer time
available to us.

The SOS model is defined on a square lattice with total energy 
\begin{equation}
{\cal H}=\sum_{\langle ij\rangle }\epsilon |h_i-h_j|
\end{equation}
where $h_i$ is the column height and $\langle ij\rangle $ denotes
nearest-neighbor pairs on a square lattice. Surface diffusion is simulated
by exchange of atoms on top of a nearest-neighbor pair of columns ($%
h_i\rightarrow h_i-1$ and $h_j\rightarrow h_j+1$, where $i,j$ are
nearest-neighbor sites). A driving force is simulated by asymmetric attempt
frequencies in directions along and opposite the force direction. For
example, when the force is in $x$ direction, we assume the attempt
frequencies in the $+x$ and $-x$ directions satisfy the following relation: 
\begin{equation}
p_{+x}/p_{-x}=\exp (2Fa/k_BT).
\end{equation}

Our next task is to describe how the probability $\Gamma _{ij}$ for an
adatom hopping from $i$ site to $j$ depends on surface configurations. Often
in statistical physics, {\em Metropolis dynamics} is used, where the hopping
probability depends on the energy difference between the final and initial
state $\Delta E_{ij}$ through the relation 
\begin{equation}
\Gamma _{ij}=\min [1,\exp (-\Delta E_{ij}/kT)].
\end{equation}
This dynamics often has the virtue of fast equilibration in the
absence of a driving force since there are no barriers for movements
with no change in energy, but it does not usually provide a
physically realistic description of actual dynamical processes.

Krug and Dobbs \cite {krug,krug95,dobbs96} have studied in detail the
effects of an external driving force combined with Metropolis dynamics
in a SOS model. They used these simulations along with a continuum
model to ``describe the {\em universal} features'' of the
electromigration problem. However, the resulting surface structures
have several artificial features that do not resemble experiments on
Si(111) surfaces. For example, in their simulations, a surface
instability develops regardless of the current direction and then
there are no extended flat regions of the surface with $\nabla h=0$.
Experiments on Si(111) generally reveal flat terraces and individual
steps coalescing into bunches when the current is in the unstable
direction, and reversing the current direction will {\em stabilize} the uniform
step train.  

Of course it is possible that at much later times some limiting
features of both the experiments and simulations could be insensitive
to the choice of dynamics, and hence universal. For example, most
driven surfaces eventually become ``rough''; because of transverse
step fluctuations \cite{rost96} this probably holds true in principle for the
experiments at sufficiently large length and time scales even when the
current is in the nominally ``stable'' direction. We show here that
with a more physically motivated choice of dynamics, the SOS model can
provide a qualitatively accurate description of the length and time
scales probed by present experiments, as well as of any longer time
``universal'' features, if such exist.

Diffusion on surfaces is usually an activated process with an energy
barrier. A different dynamical scheme, {\em Arrhenius dynamics, }takes this
physics into account in an extreme way by assuming that the energy barrier
is simply the binding energy of the atom, independent of the final
configuration. However, Krug {\em et al.} \cite{krug95} found that there is 
{\em no} morphological change for this dynamics under an external driving
force. They showed in general that instabilities in a continuum model are
associated with the dependence of the adatom mobility on the local slope,
while instabilities in a microscopic model require a dependence of the
hopping probability on the final configuration. Here we obtain such a
configuration dependence by modifying the original Arrhenius dynamics, which
provides a reasonable description of activated processes such as surface
diffusion, to include an {\em extra barrier} that arises from the presence
of steps.

This energy barrier is motivated by the physics of rebonding and surface
reconstruction that can occur near steps. The surface atoms near steps on
Si(111) surfaces usually rearrange themselves and rebond in characteristic
ways to lower the step energy.\cite{kodiyalam95} To incorporate an {\em %
additional} adatom into the step usually involves the collective motion of
many atoms as this rebonding is modified. This process has a higher
activation energy than the simple pairwise additive bond picture in the
usual SOS model would suggest. Also, in many cases the repeatable step unit,
the kink, has a complex structure, and requires the incorporation of two
adatoms to bring about its movement. To take account of this physics in our
simulations in a simple way, we assign an additional barrier for any
movement that lowers the energy, since all attachment events are associated
with a decrease in energy. So that detailed balance holds in equilibrium,
the same barrier must also be added to a movement that increases the energy.
We call this scheme {\em modified Arrhenius dynamics} and thus assume 
\begin{equation}
\Gamma _{ij}=\left\{ 
\begin{array}{ll}
\exp (-2\,\epsilon \,n_i),\quad & \Delta E_{ij}=0 \\ 
b\,\exp (-2\,\epsilon \,n_i),\quad & \Delta E_{ij}\neq 0,
\end{array}
\right.
\end{equation}
where $b<1$ and $n_i$ is the number of horizontal bonds the surface atom at
site $i$ has.

This way of introducing an attachment barrier was suggested by Bartelt {\em %
et al.} \cite{bartelt2} in their study of step fluctuations. We view
modified Arrhenius dynamics as a convenient but not necessarily unique
microscopic scheme that produces the ``right'' boundary conditions (giving
in particular a finite value for the kinetic coefficient $\kappa $) in the
mesoscopic step models discussed in previous sections. Thus the dynamical
behavior of mesoscopic and macroscopic scale features in the simulations
should be physically meaningful.

\begin{figure}[tb]
\centerline{\psfig{file=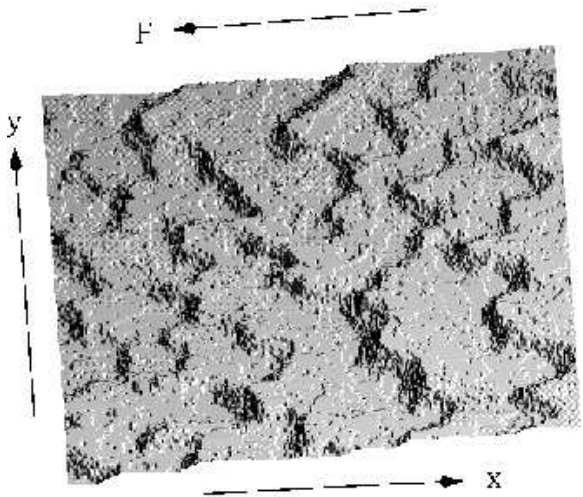,width=3.in}}
\caption{Snapshot of a simulation using a solid-on-solid model with a
diffusion bias perpendicular to the step edge direction, after about
$2.1\times 10^6$ Monte-Carlo steps. Parameters used here are:
$kT=0.8\epsilon $, $p_x=0.3$, $ p_{-x}=0.7$, $p_y=p_{-y}=0.5$, and the
size of the system is $512\times 512$ .  The dark lines are normal (up)
steps and white lines are antisteps (down steps).}
\label{sos21biasb1a}
\end{figure}

We start the simulations with a uniform step train with steps orientated
along the vertical ($y$) direction. The height of the surface increases
along the positive $x$ direction. Periodic boundary conditions are used
along the $y$ direction. In the $x$ direction we require $%
h(x+L_x,y)=h(x,y)+N_0$, where $N_0$ is the initial number of steps in the
system. For a system of size $L_x\times L_y$, the initial average terrace
width $w_0=L_x/N_0$. With a diffusion bias in the average step down
direction ($p_{+x}<p_{-x},p_{+y}=p_{-y}$), the system is unstable towards
step bunching. The step bunches continue to coarsen, consistent with the
results of previous sections. Figure \ref{sos21biasb1a} is a snapshot of a
typical configuration after some bunching has occurred. The dark regions are
step bunches, and single height crossing steps are visible between them.

The qualitative features of the simulations are very similar to the
experiments, and also to the predictions of the step model. Vicinal
surfaces are stable during the time simulated when the driving force
is in the step up direction, and unstable towards step bunching when
the force is in the step down direction. Crossing steps form when
there is significant evaporation. Preliminary results show that the
coarsening rate is consistent with the $t^{1/2}$ power law, but so far
the system size and simulation time are too small to determine the
exponent accurately.

In the 2D step model studied in Ref.~\onlinecite{kandel95} and 
\onlinecite{weeks97}, the steps are all ascending (or descending) at a given 
$y$ position. Although there is significant step bending, steps cannot form
overhangs since step positions $x_n(y)$ are defined as single-valued
functions of $y$. In the SOS model, there is no such restriction. Indeed, we
can see from Fig.~\ref{sos21biasb1a} that some crossing steps have bent so
much that they have created {\em anti-steps} at certain $y$ positions, i.e.,
steps of opposite sign to the initial ones at particular fixed $y$
positions. In our simulations, the temperature is still well below the
roughening temperature of the flat surface, but it is not energetically
forbidden to create new steps or overhangs, in contrast to the step models
previously studied.

\begin{figure}[tbp]
\centerline{\psfig{file=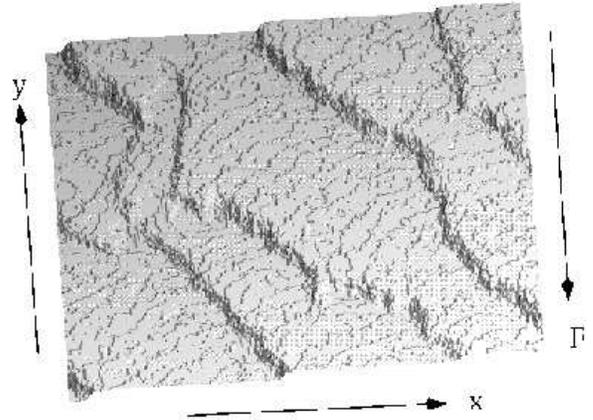,width=3.in}}
\caption{The same parameters as in Fig. \ref{sos21biasb1a}, except
that the driving force direction is parallel to the average step edge
direction ($p_y=0.3,p_{-y}=0.7,p_x=p_{-x}=0.5$).  The initial steps
are along the vertical ($y$) direction. }
\label{sos16biasb2}
\end{figure}

It is interesting to compare these results with the experiment by Latyshev 
{\em et al.}\cite{latyshev94} They observed {\em anti-step bunch formation}
taking place after step bunch formation. The first stage of the anti-step
bunch formation occurs through the bending of the single height crossing
steps between the step bunches, creating a region of bunched steps of the
opposite sign. Indeed, we have directly observed step bunches created from
this kind of step bending in our model with modified Arrhenius dynamics when
we applied the external force in a direction {\em parallel} to the initial
(and average) step edge direction. In Fig. \ref{sos16biasb2}, the initial
(and average) step edge direction is in the vertical ($y$) direction. The
bias is in the downwards ($-y$) direction. The dark regions are step bunches
formed by steps bending in the opposite direction to those individual steps
(black lines) on the terraces. As in the previous case, following the bias ($%
-y$) direction, there are regions of steps going up, and regions of
anti-step bunches going down. We derive elsewhere \cite{liu98a} from a 2D
BCF-like model a new linear instability when the diffusion bias is parallel
to the step edge direction that we believe underlies the patterns seen here.

These features obtained from simulations of the new SOS model are quite
different from the ripple structure reported by Dobbs and Krug \cite{dobbs96}
using Metropolis dynamics, where there are no distinct steps and facets
after the surface develops large structures. Here the steps and terraces are
easily discernible. Because of our more realistic treatment of the physics
of surface diffusion and attachment at steps and the favorable comparison
with experiment, we believe that {\em modified Arrhenius dynamics} provides
a better description for current-induced step bunching on Si(111).

\section{Conclusion}

\label{sec:conclusion}

In summary, the evolution of the structure of Si(111) surfaces during
electromigration at 900$^\circ$ can be understood quantitatively using a
one-dimensional step model, with parameter values and the mass transport
mechanism determined from experiment. Specifically, the $t^{1/2}$ power law
growth rate for step bunch sizes is reproduced. We concentrate on the case
where mass transport is limited by the rate of adatom attachment to a step
edge, but the method can be easily generalized, as illustrated by our
discussion of direct adatom hopping between different terraces.

The 1D step model has averaged over the individual movement of adatoms and
atomic scale fluctuations of the steps, thus permitting simulations of the
long time behavior using realistic parameter values. However, at higher
temperatures, when 2D effects such as step bending can be seen, even the
quasi-2D step models considered to date \cite{kandel95,weeks97} may not be
sufficient. Moreover an exact BCF-like treatment of full 2D diffusion
problem seems prohibitively difficult. To examine these issues, we carried
out Monte-Carlo simulations of a 2D solid-on-solid model, using physically
motivated assumptions about the dynamics of surface diffusion and attachment
at step edges. In particular we used {\em modified Arrhenius dynamics} with
an extra barrier for attachment of adatoms at step edges and find good
qualitative agreement with experiment. A new step bending instability is
seen when there is a force acting on adatoms along the step edge direction
that may be related to experiments by Latyshev {\em et al.}\cite{latyshev94}
In general we believe that this approach of combining information from
experiment, microscopic simulations, and mesoscopic step models may prove
useful in a number of different problems in surface science.

\acknowledgements

We are grateful to D. Kandel, H.-C. Jeong, O. Pierre-Louis, and E. D.
Williams for helpful discussions. This work has been supported by the
National Science Foundation (NSF-MRSEC grant \#DMR96-32521) and by grant No.
95-00268 from the United States-Israel Binational Science Foundation (BSF),
Jerusalem, Israel.

\appendix

\section{}

By requiring $j_n^{+}=j_n=j_{n+1}^{-}$ and imposing periodic boundary
conditions in Eqs.~(\ref{permplus}) and (\ref{permminus}), we arrive at the
following system of linear equations for the terrace chemical potentials for
a system of $N$ steps: 
\begin{eqnarray}
\left[ 
\begin{array}{ccccc}
c_0 & c_1 & 0 & \cdots & c_1 \\ 
c_1 & c_0 & c_1 & \cdots & 0 \\ 
0 & c_1 & c_0 & \cdots & 0 \\ 
\vdots & \vdots & \vdots & \ddots & \vdots \\ 
c_1 & 0 & 0 & \cdots & c_0
\end{array}
\right] \left[ 
\begin{array}{c}
\mu _1^t \\ 
\mu _2^t \\ 
\mu _3^t \\ 
\vdots \\ 
\mu _N^t
\end{array}
\right] =\kappa \left[ 
\begin{array}{c}
\mu _1+\mu _2 \\ 
\mu _2+\mu _3 \\ 
\mu _3+\mu _4 \\ 
\vdots \\ 
\mu _N+\mu _1
\end{array}
\right]
\end{eqnarray}
where 
\begin{equation}
c_0=2(\kappa +p)\quad c_1=-p,
\end{equation}
and $\mu _i^t=\mu ^t(\frac{x_i+x_{i+1}}2)$. $\mu _i^t$ can be solved for
analytically since the matrix on the LHS of the equation is a circulant \cite
{bellman60} matrix. The result can be expressed as 
\begin{equation}
\mu _n^t=\sum_mK_m(\mu _{n+m}+\mu _{n+m+1}),
\end{equation}
where in the limit $N\rightarrow \infty $ 
\begin{equation}
K_m=\frac \kappa {2(\kappa +p)}\frac 1{\sqrt{1-a^2}}\left( \frac{1-\sqrt{%
1-a^2}}a\right) ^m,
\end{equation}
and 
\begin{equation}
a=\frac p{\kappa +p}.
\end{equation}
$K_m$ describes the correlation between the adatom chemical potential at a
given terrace with the adatom chemical potential $m$ steps away. $K_m$
decays exponentially as $m$ increases. It is convenient to define $N_c$ as
the number of steps over which $K_m$ decreases by half, i.e., 
\begin{equation}
K_{N_c}=K_0/2.
\end{equation}
When $p\gg \kappa $, we have 
\begin{equation}
N_c\approx \log (2)\sqrt{\frac p{2\kappa }}.
\end{equation}
Since $N_c$ is the range of correlation between the chemical potential
values on different terraces, the mass transport is effectively non-local
over a number of steps much smaller than $N_c$, and is local over a number
of steps much larger than $N_c$.

\bibliographystyle{prsty}
\bibliography{bib}

\begin{thebibliography}{10}

\bibitem{latyshev}
A.~V. Latyshev, A.~L. Aseev, A.~B. Krasilnikov, and S.~I. Stenin, Surf. Sci.
  {\bf 213},  157  (1989).

\bibitem{stoyanov}
S. Stoyanov, Jpn. J. Appl. Phys. {\bf 30},  1  (1991).

\bibitem{kandel96}
D. Kandel and E. Kaxiras, Phys. Rev. Lett. {\bf 76},  1114  (1996).

\bibitem{misbah}
C. Misbah, O. Pierre-Louis, and A. Pimpinelli, Phys. Rev. B {\bf 51},  17283
  (1995).

\bibitem{kandel95}
D. Kandel and J.~D. Weeks, Phys. Rev. Lett. {\bf 74},  3632  (1995).

\bibitem{sato}
M. Sato and M. Uwaha, Phys. Rev. B {\bf 51},  11172  (1995).

\bibitem{natori}
A. Natori, Jpn. J. Appl. Phys. {\bf 33},  3538  (1994).

\bibitem{misbah96}
C. Misbah and O. Pierre-Louis, Phys. Rev. B {\bf 53},  R4318  (1996).

\bibitem{krug}
J. Krug and H.~T. Dobbs, Phys. Rev. Lett. {\bf 73},  1947  (1994).

\bibitem{krug95}
J. Krug, H.~T. Dobbs, and S. Majaniemi, Z. Phys. B {\bf 97},  281  (1995).

\bibitem{dobbs96}
H. Dobbs and J. Krug, J. Phys. I France {\bf 6},  413  (1996).

\bibitem{williams4}
E.~D. Williams, E. Fu, Y.-N. Yang, D. Kandel, and J.~D. Weeks, Surf. Sci. {\bf
  336},  L746  (1995).

\bibitem{yang1}
Y.-N. Yang, E.~S. Fu, and E.~D. Williams, Surf. Sci. {\bf 356},  101  (1996).

\bibitem{fu}
E.~S. Fu, M.~D. Johnson, D.-J. Liu, J.~D. Weeks, and E.~D. Williams, Phys. Rev.
  Lett. {\bf 77},  1091  (1996).

\bibitem{fu97}
E.~S. Fu, D.-J. Liu, M.~D. Johnson, J.~D. Weeks, and E.~D. Williams, Surf. Sci.
  {\bf 385},  259  (1997).

\bibitem{chalmers}
S.~A. Chalmers, J.~Y. Tsao, and A.~C. Gossard, J. Appl. Phys. {\bf 73},  7351
  (1993).

\bibitem{tanaka97}
S. Tanaka, N.~C. Bartelt, and J.~M. Blakely, Phys. Rev. Lett. {\bf 78},  3342
  (1997).

\bibitem{burton}
W.~K. Burton, N. Cabrera, and F.~C. Frank, Proc. R. Soc. London, Ser. A {\bf
  243},  299  (1951).

\bibitem{rettori}
A. Rettori and J. Villain, J. Phys. (Paris) {\bf 49},  257  (1988).

\bibitem{gruber}
E.~E. Gruber and W.~W. Mullins, J. Phys. Chem. Solids {\bf 28},  875  (1967).

\bibitem{nozieres1a}
See for example, P. Nozi\`eres, in {\it Solids far from equilibrium}, edited by
  C. Godr\`eche (Cambridge University Press, Cambridge, 1992), p. 99.

\bibitem{liu96}
D.-J. Liu, E.~S. Fu, M.~D. Johnson, J.~D. Weeks, and E. Williams, J. Vac. Sci.
  Technol. B {\bf 14},  2799  (1996).

\bibitem{liu97a}
D.-J. Liu, J.~D. Weeks, and D. Kandel, Surf. Rev. Lett. {\bf 4},  107  (1997).

\bibitem{weeks97}
J.~D. Weeks, D.-J. Liu, and H.-C. Jeong,  in {\em Dynamics of Crystal Surfaces
  and Interface}, {\em Fundamental Materials Research}, edited by P.~M. Duxbury
  and T.~J. Pence (Plenum Press, New York and London, 1997), p.\ 199.

\bibitem{note}
See for example, Refs. 2 and 8. Also B. Houchmandzade and C. Misbah and A.
  Pimpinelli, J. Phys. I, France, {\bf 4}, 1843 (1994).

\bibitem{bartelt}
N.~C. Bartelt, J.~L. Goldberg, T.~L. Einstein, E.~D. Williams, J.~C. Heyraud,
  and J.~J. M\'{e}tois, Phys. Rev. B {\bf 48},  15453  (1993).

\bibitem{pimpinelli1}
A. Pimpinelli, J. Villain, D.~E. Wolf, and J.~J. M\'{e}tois, Surf. Sci. {\bf
  295},  371  (1993).

\bibitem{quasistatic}
The quasistatic approximation is well justified in our case. For example, the
  typical time for an adatom to diffuse a distance of the average terrace width
  is about $10^{-4} {\rm s}$, but the typical time for a step to travel the
  average terrace width is about $10^2$s.

\bibitem{alfonso}
C. Alfonso, J. Bermond, J. Heyraud, and J. M\'{e}tois, Surf. Sci. {\bf 262},
  371  (1992).

\bibitem{williams2}
E.~D. Williams, Surf. Sci. {\bf 299/300},  502  (1994).

\bibitem{schwoebel69}
R.~L. Schwoebel, J. Appl. Phys. {\bf 40},  614  (1969).

\bibitem{tersoff}
J. Tersoff, Z.~Z. Y.~H.~Phang, and M.~G. Lagally, Phys. Rev. Lett. {\bf 75},
  2730  (1995).

\bibitem{nozieres}
P. Nozi\`eres, J. Phys. (Paris) {\bf 48},  1605  (1987).

\bibitem{stoyanov2}
S. Stoyanov, Surf. Sci. {\bf 370},  345  (1997).

\bibitem{latyshev94}
A.~V. Latyshev, A.~B. Krasilnikov, and A.~L. Aseev, Surf. Sci. {\bf 311},  395
  (1994).

\bibitem{rost96}
M. Rost, P. \u{S}milauer, and J. Krug, Surf. Sci. {\bf 369},  393  (1996).

\bibitem{kodiyalam95}
S. Kodiyalam, K.~E. Khor, and S.~D. Sarma, Phys. Rev. B {\bf 51},  5200
  (1995).

\bibitem{bartelt2}
N.~C. Bartelt, T.~L. Einstein, and E.~D. Wiiliams, Surf. Sci. {\bf 312},  411
  (1994).

\bibitem{liu98a}
D.-J. Liu, J.~D. Weeks, and D. Kandel, in preparation. (unpublished).

\bibitem{bellman60}
R. Bellman,  in {\em Introduction to matrix analysis} (McGraw-Hill Book
  company, Inc, New York, 1960), Chap.~12, p.\ 234.

\end{thebibliography}

\end{document}